# Entangled Singularity patterns of Photons in Ince-Gauss modes

Mario Krenn[1,2,*], Robert Fickler[1,2], Marcus Huber[3], Radek Lapkiewicz[1,2], William Plick[1,2], Sven Ramelow[1,2] & Anton Zeilinger[1,2,4,*]

[1] Quantum Optics, Quantum Nanophysics, Quantum Information, University of Vienna, Boltzmanngasse 5, Vienna A-1090, Austria

[2] Institute for Quantum Optics and Quantum Information, Boltzmanngasse 3, Vienna A-1090, Austria

[3] University of Bristol, Department of Mathematics, Bristol BS8 1TW, U.K.

[4] Vienna Center for Quantum Science and Technology, Faculty of Physics, University of Vienna, Boltzmanngasse 5, Vienna A-1090, Austria

* Correspondence to M.K. (mario.krenn@univie.ac.at) and A.Z. (anton.zeilinger@univie.ac.at), telephone: 0043 1 4277 29571

**Abstract**

Photons with complex spatial mode structures open up possibilities for new fundamental high-dimensional quantum experiments and for novel quantum information tasks. Here we show for the first time entanglement of photons with complex vortex and singularity patterns called Ince-Gauss modes. In these modes, the position and number of singularities vary depending on the mode parameters. We verify 2-dimensional and 3-dimensional entanglement of Ince-Gauss modes. By measuring one photon and thereby defining its singularity pattern, we non-locally steer the singularity structure of its entangled partner, while the initial singularity structure of the photons is undefined. In addition we measure an Ince-Gauss specific quantum-correlation function with possible use in future quantum communication protocols.

# Introduction

Quantum entanglement is one of the most significant features of quantum mechanics. It is used in quantum information protocols for quantum cryptography, quantum teleportation and quantum computation [1]. Qubit entanglement of photons has been shown in various degrees of freedom, such as polarization [2], time and energy [3], path [4] or frequency [5]. In addition, photons can also be entangled in orbital angular momentum (OAM) states [6]. These so-called Laguerre-Gauss (LG) modes define a discrete, infinite-dimensional Hilbert space [7,8] and have been used in fundamental experiments concerning higher-dimensional entanglement [9-11] and cryptography [12], two-dimensional entanglement of high quanta of OAM [13,14], entanglement of 3-dimensional structures [15], as well as quantum communication in free-space [16,17]. The $LG_{n,l}$ modes are described by two quantum numbers n and l. They have (n+1) intensity rings [18] and a central phase vortex with one singularity of order l [19]. Singularities, the centres of phase vortices, are points where the phase is undefined. Their order corresponds to the topological charge of the mode.

Here we focus on modes that can have very complex vortex and singularity patterns – the so-called Ince-Gauss (IG) modes [20,21], which are a natural generalization of LG modes in elliptic coordinates. In addition to the two quantum numbers of LG modes, they have one additional continuous parameter: the ellipticity [22]. Each value of the ellipticity defines a different complete orthonormal basis set, the LG modes emerge as a special case for $\varepsilon=0$. The ellipticity leads to several unique phenomena, such as the splitting of the singularity of LG modes with topological charge $l$ into $l$ separate singularities each with unit topological charge, and the formation of additional singularities in the outer rings [23,24]. The number of singularities can be defined by choosing the topological charge and their positions can be adjusted by varying the ellipticity. Here we present the first quantum experiment with Ince-Gauss modes [25]. It is well-known from the famous EPR gedankenexperiment [26] that a measurement on one particle immediately defines the state of its entangled partner. In our experiment such a measurement of a specific singularity pattern on one side defines the singularity structure on the distant photon, while such a singularity structure was not an element of reality before.

In the following we describe experiments in which we measure IG-qubit coincidence-fringes and use a 2-dimensional entanglement-witness and a steering-inequality to verify entanglement and the non-local steering of complex singularity patterns. Then we record the coincidences for the same mode numbers, but with different ellipticities $\varepsilon$. This effect is unique for IG modes, and might be used in novel quantum information protocols. In the end, we introduce a new method to prove entanglement based on a 3-dimensional entanglement-witness, and therewith verify that the produced state is entangled in a higher-dimensional Hilbert space.

## Ince-Gauss modes

Ince-Gauss modes are the natural solutions of the Paraxial Wave Equation in elliptical coordinate system. The 2-dimensional elliptical coordinate system is described by the radial and angular elliptic coordinate u and v. In the waist plane z=0 the transformation between elliptical (u, v) and Cartesian (x, y) coordinates is given by

$$\begin{pmatrix} x \\ y \end{pmatrix} = f_0 \begin{pmatrix} \cosh(u)\cos(v) \\ \sinh(u)\sin(v) \end{pmatrix}. \qquad (1)$$

$f_0$ is the semi-focal separation (eccentricity) of the coordinate system. A separation ansatz is used to solve the Paraxial Wave Equation in elliptical coordinates [27,28]. This leads to the Ince equation which can be solved by the Ince polynomials, and gives the even and odd Ince-Gauss modes [20]

$$IG^e_{p,m,\varepsilon}(\vec{r}) = N_e \cdot C^m_p(iu,\varepsilon) \cdot C^m_p(v,\varepsilon) \cdot \exp\left(-\frac{r^2}{\omega_0^2}\right) \qquad (2)$$

$$IG^o_{p,m,\varepsilon}(\vec{r}) = N_o \cdot S^m_p(iu,\varepsilon) \cdot S^m_p(v,\varepsilon) \cdot \exp\left(-\frac{r^2}{\omega_0^2}\right) \qquad (3)$$

$\varepsilon = \frac{2f_0}{\omega_0^2}$ is the ellipticity parameter, $\omega_0$ is the beam radius at the waist, p and m are the IG mode indices with integer values. For equal ellipticity, modes with different p or m are orthogonal. As p and m both can take any positive integer value, they define an infinite-dimensional Hilbert space. The expressions $C^m_p(u,\varepsilon)$ and $S^m_p(u,\varepsilon)$ are the even and odd Ince polynomials, $N_e$ and $N_o$ are normalisation constants. Then the helical Ince-Gauss (further referred to as Ince-Gauss) modes can be defined as superpositions of even and odd Ince-Gauss modes [29]

$$IG^\pm_{p,m,\varepsilon}(\vec{r}) = \frac{1}{\sqrt{2}}\left(IG^e_{p,m,\varepsilon}(\vec{r}) \pm i \cdot IG^o_{p,m,\varepsilon}(\vec{r})\right) \qquad (4)$$

Taking the limit $\varepsilon \to 0$, Ince-Gauss modes become Laguerre-Gauss modes with an integer OAM value l=m, with the central singularities moving to the centre of the beam. In the limit of $\varepsilon \to \infty$, the Ince-Gauss modes become "helical" Hermite-Gauss modes [30]. This transition can be seen in Figure 1.

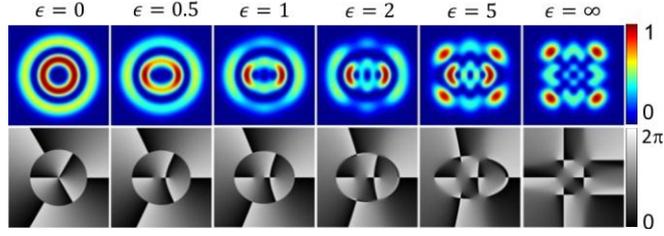

Figure 1: Ince-Gauss modes of varying ellipticity. The mode $IG_{5,3,\varepsilon}$ has two rings and three central singularities. In the upper/lower row the intensity/phase distribution of each mode is shown. From left to right the values of $\varepsilon$ are [0; 0.5; 1.0; 2.0; 5.0; $\infty$]. The splitting of the phase singularity in the centre into three singularities on a horizontal line can be observed as well as the creation of additional pairs of singularities in the ring of zero intensity [23]. For $\varepsilon=0$, the IG beam becomes a Laguerre-Gauss mode with continuous rotational symmetry of the intensity pattern. For $\varepsilon > 0$, only a 2-fold rotational symmetry remains. In the limit of $\varepsilon \rightarrow \infty$, a 4-fold rotational symmetry emerges; the corresponding modes are called helical Hermit-Gauss modes [30].

## Experiments

In our experimental setup (Figure 2), we employ type-II spontaneous parametric down-conversion (SPDC) in a nonlinear crystal (periodically poled potassium titanyl phosphate, ppKTP) which creates pairs of photons. The two photons are collinear and have orthogonal polarizations. We split the photons with a polarizing beam splitter. In the two arms of the setup we analyse by using a combination of Spatial Light Modulators (SLMs) and single mode fibres (SMFs). An SLM is a liquid crystal display, which can perform an arbitrary phase transformation on the incoming beam. In our experiment we use computer-generated holograms to convert specific higher order modes into a Gauss mode, which we couple into a SMF. Since the SMFs only allow coupling of Gauss modes, we thereby realise a spatial-mode specific filter. The photons are then detected with single-photon detectors and pairs are counted using a coincidence-logic.

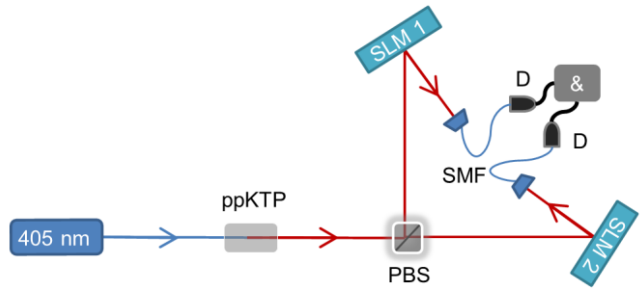

Figure 2: Schematic sketch of the experimental setup. We pump a 5mm nonlinear ppKTP crystal with a 405nm 60mW single-mode diode laser, and obtain 810nm down-converted

spatially entangled photons of orthogonal polarization. We separate the two photons on a polarizing beam splitter (PBS), and manipulate their spatial mode using Spatial Light Modulators (SLM), which transform specific Ince-Gauss modes into Gauss modes. The photons in the Gauss modes are then filtered by coupling into single mode fibres (SMF). Finally, they are detected with avalanche photo-diodes (D) and analysed with a coincidence-logic (&) with around 5ns coincidence window. The SLMs are in the far field of the crystal, and the SMFs are in the far field of the SLMs.

**2-dimensional entanglement**

In the first experiment, we restrict ourselves to a 2-dimensional Hilbert space, where we define a Bloch sphere analogously to the one representing the polarization of photons (Figure 3). The poles are helical IG modes; each point on the equator represents a specific superposition with a well-defined phase. The whole Bloch sphere can be represented by

$$IG_{p,m,\varepsilon}^{a,\varphi}(\vec{r}) = \left( \sqrt{a} \cdot \exp(i\varphi) \cdot IG_{p,m,\varepsilon}^{+}(\vec{r}) + \sqrt{1-a} \cdot \exp(-i\varphi) \cdot IG_{p,m,\varepsilon}^{-}(\vec{r}) \right), \tag{5}$$

where $a$ goes from 0 to 1, and $\varphi$ goes from 0 to $\pi$.

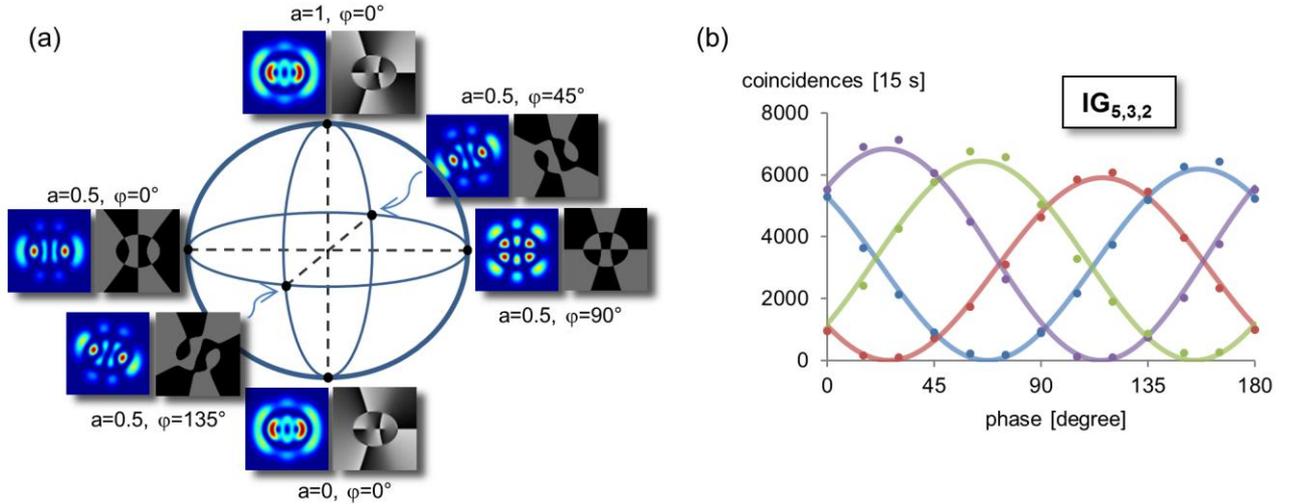

Figure 3: a: Bloch sphere constructed from the Ince-Gauss modes $IG_{5,3,2}$ with two rings and three phase singularities. The insets show the intensity (left) and phase patterns (right). Similarly to the Laguerre-Gauss (LG) modes, the intensity patterns at the North and South poles are identical. However, in contrast to LG modes, where a continuous phase change of $\varphi$ along the equator only leads to a rotation of the phase and intensity pattern, these patterns also change their shape continuously. The consequence is a different decomposition into LG modes, which leads to additional effects for non-maximal entangled LG states such as those from down-conversion. b: Coincidence fringes for the $IG_{5,3,2}$ mode with four different settings for the signal photon (22.5°, 67.5°, 112.5° and 157.5°, respectively) and 15° phase steps of the superposition for the idler, with $\sin^2$-fits. Each point has been measured for 15 seconds. We estimate the statistical uncertainty assuming a Poisson distribution of the count rates and obtained error bars are smaller than the symbols in the figure.

As a specific example, we analyse the $IG_{5,3,2}^{a,\varphi}$ mode, which has two rings and three split singularities with an ellipticity $\varepsilon = 2$ (Figure 3a). On both SLMs we display the phase-pattern for states at the equator of the Bloch sphere. The hologram for four specific phases φ is displayed at the SLM1, while the SLM2 scans through the holograms for phases from φ=0 to φ=180°. In Figure 3b and Figure 4 the coincidence counts are shown as a function of the phase of the hologram displayed at SLM2. We observe non-classical two-photon fringes, with a high visibility.

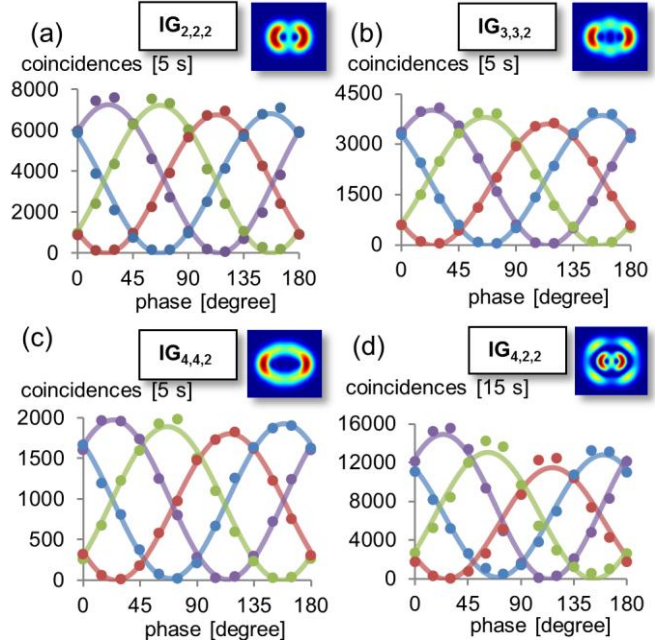

Figure 4: Coincidence fringes for different Ince-Gauss modes with ellipticity $\varepsilon$ =2. Figure a, b and c have 2, 3 and 4 singularities and no additional rings and each data point has been measured for 5 seconds. Figure d has two singularities and one additional ring. Each point has been measured for 15 seconds. The different fringes correspond to the measurement setting of the SLM2 (22.5°, 67.5°, 112.5° and 157.5°, respectively). The obtained error bars from Poisson distribution are smaller than the symbols in the figure. The lines show sinusoidal fits. We expect the different sinusoidal amplitudes as well as the deviation of the experimental points from the sinusoidal fits to be due to the plane-wave approximation used in programming the holograms.

For quantifying the entanglement, we take advantage of an entanglement witness operator [31]. Similar to entangled OAM states from down-conversion, we expect a Bell state close to $|\psi_+\rangle$ of the down-converted photon pair. Therefore a suitable witness operator for detecting entanglement in this state can be written as

$$\hat{W} = \frac{1}{4}\left(1 - \sigma_x \otimes \sigma_x - \sigma_y \otimes \sigma_y + \sigma_z \otimes \sigma_z\right), \tag{6}$$

where $\sigma_x$, $\sigma_y$ and $\sigma_z$ denote the single-qubit Pauli matrices for the two photons. The witness operator is defined to be positive for all separable states, and will give $\langle W \rangle = -0.5$ for the maximally entangled state. Recently it was shown that there exist systems that are entangled while the state of the distant photon cannot be steered[32]. We show the violation of a steering-inequality to verify that by measuring one photon, we can non-locally steer the singularity structure of its entangled partner. For that we use an inequality that has been derived recently [33,34], namely $S = |\sigma_x \otimes \sigma_x|^2 + |\sigma_y \otimes \sigma_y|^2 + |\sigma_z \otimes \sigma_z|^2 < 1$ that holds for all non-steerable states. The values of the entanglement witness W and the steering-value S calculated from our measurement results are given in Table I. The witness $\langle W \rangle$ for every measured mode is negative, which verifies entanglement for the generated states. Furthermore the steering-value S is bigger than one, which demonstrates that we are able to non-locally steer the singularity pattern of the distant photon.

| IG parameters | Witness $\langle W \rangle$ | Steering-value S |
|---|---|---|
| p=2, m=2, $\varepsilon$=2 | - 0.4847(3) | 2.879(2) |
| p=3, m=3, $\varepsilon$=2 | - 0.4897(3) | 2.918(2) |
| p=4, m=4, $\varepsilon$=2 | - 0.4905(4) | 2.925(3) |
| p=4, m=2, $\varepsilon$=2 | - 0.4581(7) | 2.675(5) |
| p=5, m=3, $\varepsilon$=2 | - 0.4784(7) | 2.830(5) |

Table I: Entanglement witness and steering-value for five different Ince-Gauss modes. For five different IG modes we have measured the entanglement witness as described in equation (6). The negative witness value verifies entanglement of our state. The steering-value S is above the non-steerable limit of S<1, which verifies that by measuring one photon we can non-locally steer the singularity pattern of the second photon. The statistical uncertainty given in brackets has been calculated assuming Poisson distributed statistics.

**Specific quantum correlation function**

In our second experiment we analyse the correlation between two down-converted photons when projected onto Ince-Gauss modes with the same mode numbers, but with different ellipticity. The ellipticity is a unique feature of IG modes, which does not exist for LG modes. It can be understood as a continuous non-trivial rotation parameter for the infinite-dimensional basis of the Hilbert space. Thus by analysing two modes with different ellipticities, we measure the overlap between continuously rotated basis elements.

As the basis rotation performed by the ellipticity parameter affects the whole infinite-dimensional Hilbert space, this has an interesting effect on two-dimensional subspaces. In contrast to simple 2-dimensional systems like polarisation, a mode with a specific ellipticity cannot fully be reconstructed in the corresponding 2-dimensional subspace with a different ellipticity. The projection into this basis gives a result smaller than one, and therefore the coincidences are reduced. It might be possible to use this effect for extensions to quantum cryptography protocols such as BB84 or Ekert91 [35,36], for instance by a two-step protocol where in the first step the secret ellipticity is transmitted and in the second step the IG basis is used and the second step uses IG modes with the ellipticity from step one. An eavesdropper faces the additional task of getting the correct value of the ellipticity, and might gain less information for a wrong value. A full security proof is out of the scope of this work.

In the experiment, we display on SLM1 the phase-pattern of an IG mode with a specific ellipticity, and on SLM2 we display the phase-pattern of a mode with the same characteristic numbers p and m but different $\varepsilon$. When the two $\varepsilon$ match, we measure a maximum coincidence count rate, whereas for different ellipticities $\varepsilon$ and $\varepsilon'$ the coincidence rate decreases. The decrease of the coincidence rate is bigger for higher modes, therefore we used $IG_{8,4,\varepsilon}$. The calculated overlap, where we assume a maximally entangled state, and the measured coincidence counts are shown in figure 5.

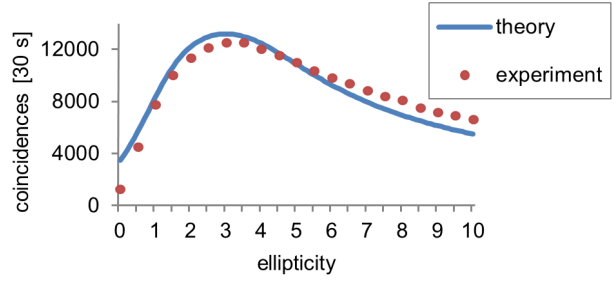

Figure 5: Overlap between two Ince-Gauss modes with the same value of p and m, but different ellipticities. This measurement shows a unique behaviour of Ince-Gauss modes, which might be useful for quantum communication applications. The horizontal axis is the ellipticity; the vertical axis shows the coincidence counts. The blue line shows the theoretical overlap $|<IG_{8,4,3}|IG_{8,4,\varepsilon}>|^2$, where the maximal overlap is at $\varepsilon=3$. The red dots are measured coincidence counts, which show good agreement to the theoretical values. The obtained error bars are smaller than the symbols in the figure.

The overlap between two Ince-Gauss modes with different ellipticities has been calculated in figure 6 for three different quantum numbers. It can be observed that the overlap drops faster for higher order modes. This can be understood when taking into account that the expansion into the Laguerre-Gauss basis involves more terms the higher the Ince-Gauss modes are. For example for IG p=14, m=6, the overlap drops very fast and reaches zero for a finite ellipticity, before it increases again. The vanishing overlap indicates orthogonal modes with the same quantum numbers (same number of initial rings and singularities). It might be interesting to investigate whether one can find multiple orthogonal modes and whether there are non-trivial relations between those orthogonal modes.

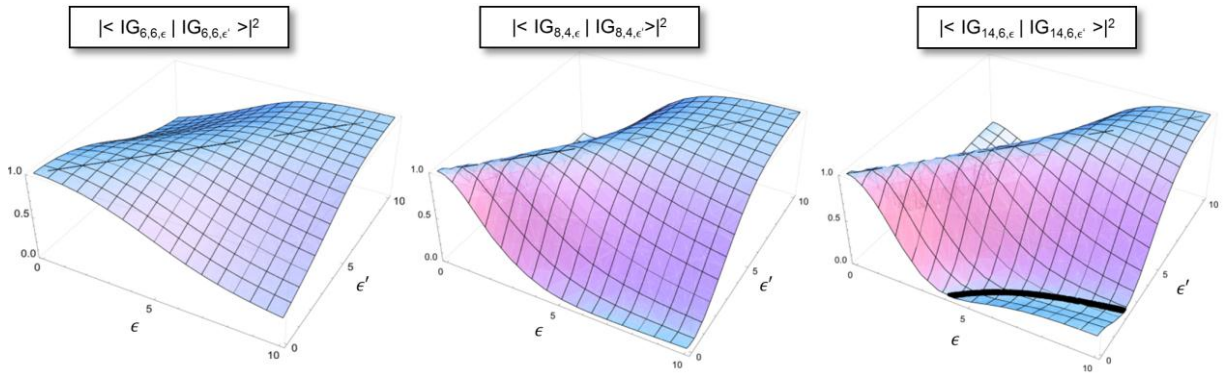

Figure 6: Overlap between Ince-Gauss modes with same quantum numbers but different ellipticity. When the ellipticities match, the overlap is maximal. When the ellipticities are different, the overlap decreases. The higher the quantum number, the faster the overlap

decreases. For p=14, m=6, the overlap reaches zero (black line) and then increases again. This vanishing overlap indicates orthogonal modes (with the same quantum number).

**Higher-dimensional entanglement**

Finally, in our third experiment, we verify that the photons are indeed entangled in a higher-dimensional Hilbert space. For this task, we use the first non-trivial 3-dimensional Ince-Gauss space. A state in such a space can be written as

$$|\psi\rangle = a\, |IG^+_{2,2,5}, IG^-_{2,2,5}\rangle + b\, |IG^+_{3,3,5}, IG^-_{3,3,5}\rangle + c\, |IG^+_{4,4,5}, IG^-_{4,4,5}\rangle, \tag{7}$$

where a, b and c are probabilities for specific modes, due to the spiral bandwidth of the SPDC process [37-39]. Similarly to the 2-dimensional case, we can define an entanglement witness for three dimensions, which consists of the visibilities in three mutually unbiased bases for every 2-dimensional subspace [40,41]. The simplest correlation function of this type can be written as

$$f(\rho) = \sum_{l=3}^{4} \sum_{k=2}^{k<l} \frac{1}{N_{k,l}} \left( \left\langle \sigma_x^{k,l} \otimes \sigma_{-x}^{k,l} \right\rangle + \left\langle \sigma_y^{k,l} \otimes \sigma_{-y}^{k,l} \right\rangle + \left\langle \sigma_z^{k,l} \otimes \sigma_{-z}^{k,l} \right\rangle \right) \tag{8}$$

The $\sigma_{\pm i}$ is a Pauli matrix constructed from $IG^\pm$ and denotes the measurements in the mutually unbiased bases of a 2-dimensional subspace of the 3-dimensional state. $N_{k,l}$ are normalisation constants that appear because we ignore the third degree of freedom in the 2-dimensional measurement.

Due to the unavoidable normalization of the 2-dimensional subspaces, bounding this correlation function for separable and 2-dimensionally entangled states becomes a challenging task. The function itself is neither linear nor convex, inevitably excluding all previously known techniques to bound such functions. We were able to prove analytical bounds for the correlation function (for details and the full analytical proof see Supplementary). The results are

- Limit for separable states: $f(\rho) = 3$
- Limit for 2-dimensionally entangled states: $f(\rho) = 6$
- Overall maximum: $f(\rho) = 9$

Inserting the measured visibilities into the correlation function in equation (8) gives the value f($\rho$)=8.156(5) which is well above the limit f($\rho$)=6 for an entangled state in two dimensions, and therefore shows that the measured state was at least a 3-dimensionally entangled state. In addition, one can use the ellipticity to tune the detection probability of higher-order spatial modes (see Figure 7), for instance to experimentally access higher-order modes than with LG modes. This might be useful for down-conversion experiments that deal with high mode numbers such as high-dimensional entanglement detection, due to the potential increase of detected count rates.

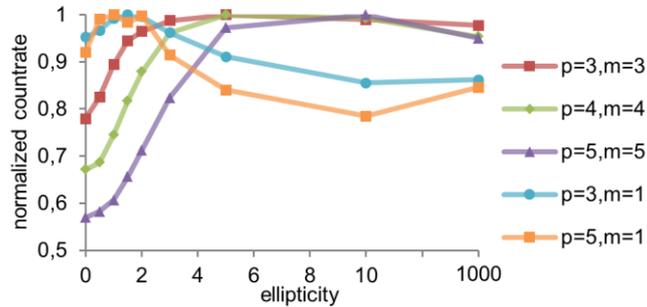

Figure 7: Normalized coincidence count-rates of several different Ince-Gauss modes depending on the ellipticity. For five different fixed p and m quantum number combination and nine values of the ellipticity, coincidence counts have been measured. For every quantum number combination, the coincidence counts have been normalized to the maximum value of the nine measurements with different ellipticity. It is shown that the probability of modes from the down-conversion can be tuned by adjusting the ellipticity of the basis. The effect is stronger for higher modes, which is crucial for high-dimensional entanglement detection. The obtained error bars are in the range of 1 percent, and therefore smaller than the symbols in the figure.

**Conclusions**

We have shown the entanglement of photons with adjustable singularity and vortex structures. By measuring a singularity pattern for the first photon we non-locally steer the positions of multiple singularities on the second photon. Additionally, we measured an IG-specific coincidence function depending on the ellipticity, which could be used for quantum information tasks. We also introduced a new method for detecting 3-dimensional entanglement, and verified therewith 3-dimensional entanglement.

We suggest that the detection probability of higher order modes from SPDC can be tuned with the ellipticity parameter, which might be useful for experiments that deal with high-order modes. Furthermore, due to the non-trivial basis rotation of the Hilbert space, we believe that this additional parameter could be used to extent well-known quantum key distribution protocols.

## Acknowledgements

This work was supported by the ERC Advanced Grant QIT4QAD, and the Austrian Science Fund FWF within the SFB F40 (FoQuS) and W1210-2 (CoQuS). M.H. gratefully acknowledges support from the EC-project "IP Q-Essence" and the ERC (Advanced Grant "IRQUAT"). We acknowledge C. Schaeff for discussion of the high-dimensional entanglement.

## References

[1] D. Bouwmeester, A. Ekert, and A. Zeilinger (eds.), Springer, (2000).

[2] S. J. Freedman, and J. F. Clauser, Phys. Rev. Lett. **28**, 938 (1972).

[3] P. G. Kwiat, A. M. Steinberg, and R. Y. Chiao, Phys. Rev. A **47**, R2472 (1993).

[4] J. G. Rarity, and P. R. Tapster, Phys. Rev. Lett. **64**, 2495 (1990).

[5] S. Ramelow, L. Ratschbacher, A. Fedrizzi, N. K. Langford, and A. Zeilinger, Phys. Rev. Lett. **103**, 253601 (2009).

[6] A. Mair, A. Vaziri, G. Weihs, and A. Zeilinger, Nature **412**, 313 (2001).

[7] G. Molina-Terriza, J. P. Torres and L. Torner, Nature Phys. **3**, 305 (2007).

[8] A. M. Yao and M. J. Padgett, Advances in Optics and Photonics **2**, 161 (2011).

[9] Vaziri A, Weihs G, and Zeilinger A., Phys. Rev. Lett. **89**, 240401 (2002).

[10] B.-J. Pors, F. Miatto, G. W. 't Hooft, E. R. Eliel, and J. P. Woerdman, Journal of Optics **13**, 064008 (2011).

[11] A. C. Dada, J. Leach, G. S. Buller, M. J. Padgett, and E. Andersson, Nature Phys. **7**, 677 (2011).

[12] S. Gröblacher, T. Jennewein, A. Vaziri, G. Weihs, and A. Zeilinger, New J. Phys. **8**, 75 (2006).

[13] J. Leach. *et al.*, Opt. Express **17**, 8287 (2009).

[14] R. Fickler *et al.*, Science **338**, 640 (2012).

[15] J. Romero *et al.*, Phys. Rev. Lett. **106**, 100407 (2011).

[16] B.-J. Pors, C. H. Monken, E. R. Eliel, and J. P. Woerdman, Opt. Express **19**, 6671 (2011).

[17] M. Malik *et al.*, Opt. Express **20**, 13195 (2012).

[18] V. D. Salakhutdinov, E. R. Eliel, and W. Löffler, Phys. Rev. Lett. **108**, 173604 (2012).

[19] L. Allen, M. W. Beijersbergen, R. J. C. Spreeuw, and J. P. Woerdman, Phys. Rev. A **45**, 8185 (1992).

[20] M. A. Bandres, and J. C. Gutiérrez-Vega, Opt. Lett. **29**, 144 (2004).

[21] M. A. Bandres, and J. C. Gutiérrez-Vega, J. Opt. Soc. Am. A **21**, 873 (2004).

[22] M. A. Bandres, and J. C. Gutiérrez-Vega, Opt. Express **16**, 21087 (2008).

[23] S. Lopez-Aguayo, and J. C. Gutiérrez-Vega, Opt. Express **15**, 18326 (2007).

[24] W. N. Plick, M. Krenn, R. Fickler, S. Ramelow, and A. Zeilinger, arXiv:1208.1865 (submitted).

[25] M. Krenn *et al.*, Bulletin of the American Physical Society **57,** BAPS.2012.MAR.W1.6, (2012).

[26] A. Einstein, B. Podolsky, and N. Rosen, Phys. Rev. **47**, 777 (1935).

[27] E. Ince, Proc. London Math. Soc. **23**, 56 (1923).

[28] F. Arscott, Pergamon Press (1964).

[29] J. B. Bentley JB, J. A. Davis, M. A. Bandres, J. C. Gutiérrez-Vega, Opt. Lett. **31**, 649 (2006).

[30] C. López-Mariscal, and J.C. Gutiérrez-Vega, Proc. of SPIE 6663 (2007).

[31] O. Gühne, and G. Tóth, Physics Reports **474**, 1 (2009).

[32] D. J. Saunders, S. J. Jones, H. M. Wiseman, G. J. Pryde, Nature Phys. **6**, 845 (2010).

[33] D. H. Smith *et al.*, Nat. Commun. **3**, 625 (2012).

[34] B. Wittmann, S. Ramelow *et al.*, New J. Phys. **14**, 053030 (2012).

[35] C. H. Bennett and G. Brassard, in Proceedings of the IEEE International Conference on Computers, Systems, and Signal Processing, Bangalore, p. 175 (1984)

[36] A. K. Ekert, Phys. Rev. Lett. **67**, 661 (1991).

[37] F. M. Miatto, A. M. Yao, and S. M. Barnett, Phys. Rev. A **83**, 033816 (2011).

[38] H. Di Lorenzo Pires, H. C. B. Florijn HCB, and M. P. van Exter, Phys. Rev. Lett. **104**, 020505 (2010).

[39] E. Brambilla, L. Caspani, L. A. Lugiato, and A. Gatti, Phys. Rev. A **82**, 013835 (2010).

[40] C. Spengler, M. Huber, A. Gabriel, and B. Hiesmayr, Quantum Information Processing 1–10 doi:10.1007/s11128-012-0369-8 (2012).

[41] J. I. de Vicente, M. Huber, Phys. Rev. A **84**, 062306 (2011).

# Bounds of the correlation function for higher-dimensional entanglement detection

This note provides the proof of the bounds for the correlation function, which we use to verify the presence of three dimensionally entangled modes. It shows how even non-convex correlation functions using two-dimensional subspaces can be bounded via a correlation tensor approach.

## PROOF THAT $f(\rho) \leq 6$ FOR TWO-DIMENSIONALLY ENTANGLED STATES

Using the following abbreviations

$$|IG^+_{2,2,5}\rangle = |0_+\rangle, |IG^+_{3,3,5}\rangle = |1_+\rangle, |IG^+_{4,4,5}\rangle = |2_+\rangle,$$
$$|IG^-_{2,2,5}\rangle = |0_-\rangle, |IG^-_{3,3,5}\rangle = |1_-\rangle, |IG^-_{4,4,5}\rangle = |2_-\rangle, \tag{S1}$$

we can represent the performed measurements via the following operators

$$\sigma^{kl}_{\pm x} := |k_\pm\rangle\langle l_\pm| + |l_\pm\rangle\langle k_\pm|,$$
$$\sigma^{kl}_{\pm y} := i|k_\pm\rangle\langle l_\pm| - i|l_\pm\rangle\langle k_\pm|,$$
$$\sigma^{kl}_{\pm z} := |k_\pm\rangle\langle k_\pm| - |l_\pm\rangle\langle l_\pm|. \tag{S2}$$

In order to lower bound the dimensionality of entanglement we use the following correlation function

$$f(\rho) = g(\rho^{01}) + g(\rho^{02}) + g(\rho^{12}), \tag{S3}$$

where $\rho^{kl}$ are the normalized subspace density matrices, where all but two degrees of freedom on both sides are ignored, i.e.

$$\rho^{kl} := \frac{(|k_+\rangle\langle k_+| + |l_+\rangle\langle l_+|) \otimes (|k_-\rangle\langle k_-| + |l_-\rangle\langle l_-|)\rho(|k_+\rangle\langle k_+| + |l_+\rangle\langle l_+|) \otimes (|k_-\rangle\langle k_-| + |l_-\rangle\langle l_-|)}{N_{kl}}, \tag{S4}$$

where $N_{kl}$ is the normalization, such that $\text{Tr}(\rho^{kl}) = 1$, and

$$g(\rho^{kl}) = \text{Tr}\left((\sigma^{kl}_{+z} \otimes \sigma^{kl}_{-z} - \sigma^{kl}_{+y} \otimes \sigma^{kl}_{-y} + \sigma^{kl}_{+x} \otimes \sigma^{kl}_{-x})\rho^{kl}\right). \tag{S5}$$

Comparing these to the correlations on the total state, i.e.

$$f_{kl} = \text{Tr}\left((\sigma^{kl}_{+z} \otimes \sigma^{kl}_{-z} - \sigma^{kl}_{+y} \otimes \sigma^{kl}_{-y} + \sigma^{kl}_{+x} \otimes \sigma^{kl}_{-x})\rho\right), \tag{S6}$$

we can write

$$g(\rho_{kl}) = \frac{f_{kl}}{N_{kl}}, \tag{S7}$$

and thus

$$f(\rho) = \sum_{k<l} \frac{f_{kl}}{N_{kl}}. \tag{S8}$$

A bound can easily be obtained for the convex function as it is maximized by pure states. We just need to optimize over pure states with Schmidt-Rank two, in order to get a bound for two dimensionally entangled states, i.e.

$$\sum_{k<l} f_{kl} \leq \max_{|\psi_2\rangle} f(|\psi_2\rangle\langle\psi_2|), \tag{S9}$$

where $|\psi_2\rangle = \lambda_1|v_1 v'_1\rangle + \lambda_2|v_2 v'_2\rangle$. Lagrangian maximization leads to three distinct maximizing states

- $|\psi_2\rangle = \frac{1}{\sqrt{2}}(|0_+0_-\rangle + |1_+1_-\rangle)$ or

- $|\psi_2\rangle = \frac{1}{\sqrt{2}}(|0_+0_-\rangle + |2_+2_-\rangle)$ or

- $|\psi_2\rangle = \frac{1}{\sqrt{2}}(|1_+1_-\rangle + |2_+2_-\rangle)$.

Then using the fact that $\Re e[\langle\alpha|\rho|\beta\rangle] \leq \frac{1}{2}(\langle\alpha|\rho|\alpha\rangle + \langle\beta|\rho|\beta\rangle)$ (due to the positivity of the density matrix) we can bound the expression via

$$\sum_{k<l} f_{kl} \leq 4(\langle 0_+0_-|\rho|0_+0_-\rangle + \langle 1_+1_-|\rho|1_+1_-\rangle + \langle 2_+2_-|\rho|2_+2_-\rangle). \tag{S10}$$

Now we are left with the following Lagrangian problem

$$f(\rho, \lambda) = \sum_{k<l} \frac{f_{kl}}{N_{kl}} + \lambda(\sum_{k<l} f_{kl} - 4(\sum_{k=0}^{2} \langle k_+k_-|\rho|k_+k_-\rangle)), \tag{S11}$$

which we want to solve in order to maximize $f(\rho)$. Taking

$$\frac{\partial}{\partial \Re e[\langle k_+k_-|\rho|l_+l_-\rangle]} f(\rho) = \frac{1}{N_{kl}} + \lambda = 0, \tag{S12}$$

shows that every extremal point has to satisfy

$$\lambda = -\frac{1}{N_{01}} = -\frac{1}{N_{02}} = -\frac{1}{N_{12}} \tag{S13}$$

$$\Rightarrow N_{01} = N_{02} = N_{12} =: N. \tag{S14}$$

In consequence for any extremal point we can directly bound

$$f(\rho) \leq \sum_{k<l} \frac{f_{kl}}{N} - \frac{1}{N}(\sum_{k<l} f_{kl} - 4(\sum_{k=0}^{2} \langle k_+k_-|\rho|k_+k_-\rangle)) = \frac{4(\sum_{k=0}^{2} \langle k_+k_-|\rho|k_+k_-\rangle)}{N}. \tag{S15}$$

Now we can use that

$$N = \frac{1}{3}(N_{01} + N_{02} + N_{12}) = \frac{2}{3}(\sum_{k=0}^{2} \langle k_+k_-|\rho|k_+k_-\rangle + \frac{\sum_{k\neq l} \langle k_+l_-|\rho|k_+l_-\rangle}{2}) \geq \frac{2}{3}(\sum_{k=0}^{2} \langle k_+k_-|\rho|k_+k_-\rangle), \tag{S16}$$

and have a lower bound on $N$, which directly leads to another lower bound on $f(\rho)$:

$$f(\rho) \leq \frac{4(\sum_{k=0}^{2} \langle k_+k_-|\rho|k_+k_-\rangle)}{\frac{2}{3}(\sum_{k=0}^{2} \langle k_+k_-|\rho|k_+k_-\rangle)} = 6. \tag{S17}$$

$\square$

Surprisingly this bound is tight, i.e. it can be saturated by the following two dimensionally entangled state

$$\rho_2 = \frac{1}{3}(|\phi_{01}^+\rangle\langle\phi_{01}^+| + |\phi_{02}^+\rangle\langle\phi_{02}^+| + |\phi_{12}^+\rangle\langle\phi_{12}^+|), \tag{S18}$$

where $|\phi_{kl}^+\rangle := \frac{1}{\sqrt{2}}(|k_+k_-\rangle + |l_+l_-\rangle)$, where $f(\rho_2) = 6$. This is intriguing as $f(|\phi_{kl}^+\rangle\langle\phi_{kl}^+|) = 5$, meaning that we have indeed bounded a non-convex entanglement detection criterion.

## PROOF THAT $f(\rho) \leq 3$ FOR SEPARABLE STATES

This follows from the fact that every $2 \times 2$ dimensional subspace of a separable state is again separable. Thus

$$g(\rho^{kl}) = \langle \sigma_{+z}^{kl} \otimes \sigma_{-z}^{kl} \rangle - \langle \sigma_{+y}^{kl} \otimes \sigma_{-y}^{kl} \rangle + \langle \sigma_{+x}^{kl} \otimes \sigma_{-x}^{kl} \rangle \tag{S19}$$

$$= \langle \sigma_{+z}^{kl} \rangle \cdot \langle \sigma_{-z}^{kl} \rangle - \langle \sigma_{+y}^{kl} \rangle \cdot \langle \sigma_{-y}^{kl} \rangle + \langle \sigma_{+x}^{kl} \rangle \cdot \langle \sigma_{-x}^{kl} \rangle \tag{S20}$$

$$= \begin{pmatrix} \langle \sigma_{+x}^{kl} \rangle \\ \langle \sigma_{+y}^{kl} \rangle \\ \langle \sigma_{+z}^{kl} \rangle \end{pmatrix} \cdot \begin{pmatrix} \langle \sigma_{-x}^{kl} \rangle \\ -\langle \sigma_{-y}^{kl} \rangle \\ \langle \sigma_{-z}^{kl} \rangle \end{pmatrix} \leq \left| \begin{pmatrix} \langle \sigma_{+x}^{kl} \rangle \\ \langle \sigma_{+y}^{kl} \rangle \\ \langle \sigma_{+z}^{kl} \rangle \end{pmatrix} \right| \cdot \left| \begin{pmatrix} \langle \sigma_{-x}^{kl} \rangle \\ -\langle \sigma_{-y}^{kl} \rangle \\ \langle \sigma_{-z}^{kl} \rangle \end{pmatrix} \right| \leq 1, \tag{S21}$$

due to the fact that the local Bloch vectors's length is limited by one. As

$$f(\rho) = \underbrace{g(\rho^{01})}_{\leq 1} + \underbrace{g(\rho^{02})}_{\leq 1} + \underbrace{g(\rho^{12})}_{\leq 1} \leq 3, \tag{S22}$$

we have completed the proof. Also this bound is tight as e.g. for $\rho_S = \frac{1}{3}(|0_+0_-\rangle\langle 0_+0_-|+|1_+1_-\rangle\langle 1_+1_-|+|2_+2_-\rangle\langle 2_+2_-|)$, it is easy to see that $f(\rho_S) = 3$.